\documentclass{amsart}
\usepackage{latexsym,amssymb,amsfonts,amsmath, graphicx, color,enumerate, url}
\usepackage{dsfont}
\usepackage[latin1]{inputenc}
\usepackage{hyperref}
\numberwithin{equation}{section}
\newtheorem{theo}{Theorem}

\newtheorem{prop}[theo]{Proposition}
\newtheorem{cor}[theo]{Corollary}
\newtheorem{defi}[theo]{Definition}

\theoremstyle{definition}

\newtheorem{rem}[theo]{Remark}
\newtheorem{rems}[theo]{Remarks}

\numberwithin{theo}{section}
 \def\mE{\mathsf{E}}

 \def\mV{\mathsf{V}}

 \def\mG{\mathsf{G}}

\title[Towards a gauge theory for evolution equations]{Towards a gauge theory for evolution equations\\ on vector-valued spaces}
\author{Stefano Cardanobile}
\address{Bernstein Center for Computational Neuroscience,  Hansastra{\ss}e 9A, D-79104 Freiburg, Germany}
\email{stefano.cardanobile@bccn.uni-freiburg.de}
\author{Delio Mugnolo}
\address{Institut f\"ur Analysis, Universit\"at Ulm, Helmholtzstra{\ss}e 18, D-89081 Ulm, Germany}
\email{delio.mugnolo@uni-ulm.de}

\thanks{The authors wish to thank Hendrik Vogt (Dresden) for helpful comments.}

\begin{document}

\begin{abstract}
We investigate symmetry properties of vector-valued diffusion and Schr\"odinger equations. For a separable Hilbert space $H$ we characterize the subspaces of $L^2({{\mathbb R}^N}, H)$ that are local (i.e., defined pointwise) and discuss the issue of their invariance under the time evolution of the differential equation. In this context, the possibility of a connection between our results and the theory of gauge symmetries in mathematical physics is explored.
\end{abstract}

\maketitle

\section{The abstract setting: global symmetries}\label{gensett}

In mathematical physics, one is often interested in the formulation of gauge theories. These are field theories where solutions of the relevant equations are symmetric -- i.e., invariant under some transformation group of the functional values. A prototypical example is given by quantum electrodynamics, which is a gauge theory with respect to the symmetry group $U(1)$ (the unitary group) and leads to the introduction of the electromagnetic field.

The usual framework to deal with gauge theories in a mathematically rigorous way is that of differential geometry. Aim of this note is to propose a possible approach based on operator theoretic methods, instead, borrowing some ideas from the theory of vector bundles. 

Let $H$ be a separable complex Hilbert space and consider the Bochner space ${\mathcal H}:=L^2({{\mathbb R}^N};H)$, which is a Hilbert space with respect to the canonical inner product
$$(f|g):=\int_{{\mathbb R}^N} (f(x)|g(x))_H dx,\qquad f,g\in L^2({{\mathbb R}^N};H).$$
Let $\mathcal V$ be a Hilbert space which is densely embedded into $\mathcal{H}$ --  typically, a vector-valued Sobolev spaces: these can be defined in a standard way, see e.g.~\cite[\S~III.4]{Ama95}. Consider an ${\mathcal H}$-elliptic, continuous, sesquilinear form $a:{\mathcal V}\times {\mathcal V}\to\mathbb C$ -- see e.g.~\cite{Ouh05} for the general theory of sesquilinear forms. Denote by $A$ the operator associated with $a$ and by $(e^{tA})_{t\geq 0}$ the semigroup generated by $A$. The form $a$ is symmetric if and only if $A$ is self-adjoint, and in this case by Stone's theorem $iA$ generates a unitary group $(e^{itA})_{t\in\mathbb R}$. We assume throughout this note that $a$ is symmetric, i.e., $a(f,g)=\overline{a(g,f)}$ for all $f,g\in \mathcal V$.

In mathematical language, a physical system is said to have a symmetry if there is a Lie group $\mathcal O$ of operators such that each $O\in \mathcal O$ commutes with the time evolution of the system. Alternatively, in the Lagrangian formulation of classical field theory, a physical system with Lagrangian functional $\mathcal L$ admits a symmetry if there exists a Lie group $\mathcal O \subset \mathcal{L}(\mathcal{H})$ such that
$${\mathcal L}(O\phi)={\mathcal L}(\phi)$$
for all $O\in\mathcal O$ and all $\phi$ smooth enough -- say, in the domain of the sesquilinear form that appears in the weak formulation of the problem.
It can be shown that these notions are equivalent in the self-adjoint case, cf. Proposition~\ref{gauge} or~\cite[\S~5]{CarMugNit08} for a more detailed treatment.

For example, let us sketch the prototypical case of the Schr\"odinger equation in ${{\mathbb R}^N}$. In this context the Lagrangian is given by
\begin{equation}\label{lagrschr}
{\mathcal L}(\phi):=\int_0^T\int_{{\mathbb R}^N} i\dot{\phi}(t,x)\overline{\phi(t,x)}+|\nabla\phi(t,x)|^2 dx dt
\end{equation}
for all $\phi\in C^1((0,T);L^2({{\mathbb R}^N}))\cap C((0,T);H^1({\mathbb R}^N))$. Introducing the (scalar-valued) complex Hilbert space ${\mathcal H}:= L^2({{\mathbb R}^N})$ and the sesquilinear form
$$a(f,g):=\int_{{\mathbb R}^N} \nabla f(x) \overline{\nabla g(x)}dx,\qquad f,g\in {\mathcal V}:=H^1({{\mathbb R}^N}),$$
(i.e., taking $H={\mathbb C}$ in the above formalism) one has
$${\mathcal L}(\phi)=\int_0^T i(\dot{\phi}(t)|{\phi(t)})_{\mathcal H}+a(\phi(t),\phi(t)) dt.$$
Thus, if $\mathcal O$ is a Lie group of unitary operators, we conclude that the system admits a symmetry if and only if
\begin{equation}\label{symmetryphys}
Of\in {\mathcal V}\qquad\hbox{and}\qquad a(Of)=a(f)\qquad\hbox{for all }f\in {\mathcal V}\hbox{ and }O\in{\mathcal O},
\end{equation}
where $a(\cdot)$ denotes the quadratic form associated with the sesquilinear form $a(\cdot,\cdot)$.
Motivated by this example, we introduce the following. 

\begin{defi}\label{symmschr}
A Lie group $\mathcal O$ of unitary operators on a Hilbert space $\mathcal H$ is said to be a \emph{global symmetry} of a Schr\"odinger equation if condition~\eqref{symmetryphys} holds.
\end{defi}

A special class of global symmetries has been introduced in~\cite{CarMugNit08}: those unitary groups whose generator is $i\mathcal P$, where $\mathcal P$ is an orthogonal projection onto some closed subspace of $\mathcal H$. We are particularly interested in a special class of orthogonal projections. To fix the ideas, let $G$ be a closed subspace of $H$ and introduce the subspace
\begin{equation}\label{projy}
\mathcal{G} := \left\{ f \in \mathcal H : f(x) \in G \mbox{ for a.\,e. } x\in {{\mathbb R}^N} \right\}.
\end{equation}
of $L^2({{\mathbb R}^N};H)$. Then $\mathcal{G}$ is closed. Therefore, we can consider the orthogonal projection $\mathcal P_{\mathcal G}$ of $L^2({{\mathbb R}^N};H)$ onto $\mathcal G$: one can easily check that this is given by
\begin{equation}\label{p2}
(\mathcal P_{\mathcal G}f)(x)=P_G(f(x))\qquad \hbox{for all }f\in L^2({{\mathbb R}^N};H)\hbox{ and a.e.\ }x\in {{\mathbb R}^N},
\end{equation}
where $P_G$ denotes the orthogonal projection of $H$ onto $G$. Then $\mathcal P_{\mathcal G}$ is self-adjoint, thus by Stone's theorem we can consider $(e^{is\mathcal P_{\mathcal G}})_{s\in\mathbb R}$, the unitary group generated by $i{\mathcal P_{\mathcal G}}$.  (In fact, since $\mathcal P_{\mathcal G}$ is bounded, it even generates an analytic group $(e^{z\mathcal P_{\mathcal G}})_{z\in\mathbb C}$).

\begin{rem}
It follows directly from  the series expansion of the exponential function that
\begin{equation}\label{explform}
e^{z\mathcal{P}} = e^{z}\mathcal{P} + (Id-{\mathcal P})=e^{z}{\mathcal P}+{\mathcal P}^\perp,\qquad z\in\mathbb C,
\end{equation} 
for any orthogonal projection $\mathcal P$ of a Hilbert space onto a closed subspace. (Here and in the following we denote by $\mathcal P^\perp$ the orthogonal projection onto ${\rm ker}\mathcal P$, i.e., $Id-{\mathcal P}$). Observe that for $z\in i\mathbb R$ it acts as $U(1)$ in the direction of ${\rm range}\mathcal P$ and as the identity in the orthogonal direction.
\end{rem}

Plugging~\eqref{p2} into~\eqref{explform} we obtain
\begin{eqnarray*}
(e^{z\mathcal P_{\mathcal G}}f)(x) &=& e^{z} (\mathcal P_{\mathcal G}f)(x) + ({\mathcal P_{\mathcal G}}^\perp f)(x)\\
& =& e^{z} P_G(f(x)) + P_G^\perp(f(x))\\
&=&e^{zP_G}(f(x))
\end{eqnarray*}
for all $z\in{\mathbb C}$, $f\in L^2({{\mathbb R}^N};H)$ and a.e.\ $x\in {{\mathbb R}^N}$. Thus, associated with $(e^{is\mathcal P_{\mathcal G}} )_{s\in\mathbb R}$ we consider $(e^{isP_G})_{s\in\mathbb R}$: clearly, also the latter group is unitary.

\begin{rem}
The latter is in fact a (compact, simply connected) \emph{Lie group} 
of dimension 1 with associated Lie algebra
$${\mathfrak g}=\{is P_G:s\in\mathbb R\},$$
and  $\Pi:e^{is P_G}\mapsto e^{is{\mathcal P_{\mathcal G}}}$ is a unitary representation. Of course, there are as many Lie groups $(e^{is\mathcal P_{\mathcal G}})_{s\in\mathbb R}$ of the above type as closed subspaces $G$ of $H$. By the general theory of Lie groups we know in particular that this representation is completely reducible, i.e., it is the direct sum of irreducible representations. In fact, by definition and due to~\eqref{explform}, the representation $\Pi$ is irreducible if and only if the only closed subspaces ${\mathcal W}$ of $L^2({{\mathbb R}^N};H)$ satisfying $\mathcal P_{\mathcal G}{\mathcal W}\subset {\mathcal W}$ are the trivial ones\footnote{ This notion of irreducibility is clearly different from that of \emph{irreducible operator} in the sense of Hilbert lattices.}. (Observe that for all closed subspaces ${\mathcal G},{\mathcal W}$ of $\mathcal H$ the three conditions $\mathcal P_{\mathcal G}{\mathcal W}\subset {\mathcal W}$, ${\mathcal P}_{\mathcal W}{\mathcal G}\subset {\mathcal G}$, and $\mathcal P_{\mathcal G}{\mathcal P}_{\mathcal W}={\mathcal P}_{\mathcal W}\mathcal P_{\mathcal G}$ are equivalent, cf.~\cite[Lemma~2.3]{ManVogVoi05}).
 
In general, one sees that a necessary condition for irreducibility of the representation $\Pi$ is that the operator $P_G$ be irreducible in the sense of Banach lattices (in the case of ${\rm dim}H<\infty$, this amounts to saying that the matrix $P_G$ cannot be placed into block triangular form by permutations of rows and columns); irreducibility of $P_G$ is not sufficient though, since not all closed subspaces $\mathcal G$ of $\mathcal H$ are of the form $\mathcal G$ for a suitable subspace $G$ of $H$. Intuitively, the representation $\Pi:e^{is P}\mapsto e^{is {\mathcal P}}$ will not in general be irreducible, as one sees already in the simple case of ${{\mathbb R}^N}=(0,1)$ and $G=H={\mathbb C}$ (i.e., $\Pi=Id$) if one considers the subspace $\mathcal W$ of radial functions with respect to the point $\frac{1}{2}$.
\end{rem}

\begin{rem}
One can also observe that by~\ref{explform} each closed subspace $G$ defines canonically a (nontrivial) vector bundle
$$S^1 \hookrightarrow S^1\times H=S^1\times G\times G^\perp\to H.$$
In particular, the bundle projection $\pi:=e^{isP_G}$ satisfies
$$\pi^{-1}(\{w\})=\pi^{-1}(\{P_G w,P_G^\perp w\})=\{(s,v)\in S^1\times G:v=e^{is}P_G w\}\times\{ P_G^\perp w\},\qquad w\in H.$$
Similarly, $G$ also defines a vector bundle $S^1 \to S^1\times \mathcal H\to \mathcal H$ with  bundle projection $\pi:=e^{is{\mathcal P}_{\mathcal G}}$. It is worth to remark that $U(1)$ is not the structure group (in the physical language, the \emph{gauge group}) of either of these bundles -- unless $G$ is trivial.
\end{rem}

One can wonder whether, given a subspace $G$ of $H$, the subspace $\mathcal G$ is invariant under $(e^{tA})_{t\geq 0}$, provided that suitable conditions are verified by $A$, the (self-adjoint) operator associated with a symmetric sesquilinear form $a$. Such invariant properties can be characterized by a simple condition. Let us recall the following result yielding a characterization of invariant subspaces. It has been proved in~\cite[\S 5]{CarMugNit08} in the case of an orthogonal projection whose range is a closed subspace defined as in~\eqref{projy}, but its proof carries over verbatim to the general case. We repeat here its proof for  the sake of self-containedness 
(and to fill a small gap in the proof in~\cite{CarMugNit08}).

\begin{prop}\label{gauge}
Let $\mathcal{P}$ be an orthogonal projection onto a closed subspace of a Hilbert space $\mathcal H$ and let $a:{\mathcal V}\times{\mathcal V}\to\mathbb C$ be an $\mathcal H$-elliptic, continuous, symmetric, densely defined sesquilinear form with associated operator $A$. Then the following assertions are equivalent.
\begin{enumerate}[(a)]
\item\label{invariance}
The range of $\mathcal P$ is invariant under $(e^{tA})_{t \ge 0}$.
\item\label{formproj}
If $\psi\in {\mathcal V}$, then ${\mathcal P}\psi \in {\mathcal V}$ and $a(\mathcal{P} \psi, \psi) = a(\mathcal{P}\psi, \mathcal{P}\psi)$.
\item\label{formgroup}
If $\psi\in {\mathcal V}$, then $e^{is\mathcal{P}}\psi\in {\mathcal V}$ and $a(\psi, \psi) = a(e^{is\mathcal{P}}\psi, e^{is\mathcal{P}}\psi)$ for all $s\in\mathbb R$.
\end{enumerate}
\end{prop}

\begin{proof}
Due to Ouhabaz's invariance criterion (cf.~\cite[Thm.~2.1]{ManVogVoi05} for a generalized version), \eqref{invariance} is equivalent to $\mathcal{P}{\mathcal V} \subset {\mathcal V}$ and $a(\mathcal{P}\psi, (Id-\mathcal{P})\psi) = 0$ for every $\psi \in {\mathcal V}$. This is precisely~\eqref{formproj}.

By~\eqref{explform},  ${\mathcal V}$ is invariant under $\mathcal{P}$ if and only if it is invariant under the action of $e^{is\mathcal{P}}$. Moreover, this formula implies
	\[
		a\left(e^{is\mathcal{P}}\psi, e^{is\mathcal{P}}\psi\right) \\
			= |e^{is} - 1|^2 a(\mathcal{P} \psi, \mathcal{P} \psi)
				+ 2 {\rm Re} (e^{is} - 1) a(\mathcal{P} \psi, \psi)
				+ a(\psi,\psi).
	\]
	On the one hand, the identity $|e^{is} - 1|^2 = 2 - 2 {\rm Re} e^{is}$ now shows
	that~\eqref{formproj} implies~\eqref{formgroup}. On the other hand, if~\eqref{formgroup}
	holds, then the above calculation implies
	\[
		|e^{is} - 1|^2 a(\mathcal{P} \psi, \mathcal{P} \psi) =
				- 2 {\rm Re} (e^{is} - 1) a(\mathcal{P} \psi, \psi)
	\]
	for every $s \in \mathbb{R}$. This is~\eqref{formproj} for $s = \pi$.
\end{proof}

\section{Space-dependent subspaces}\label{stca}

The aim of this section is to describe and characterize the subspaces of the vector-valued function space $\mathcal H=L^2({{\mathbb R}^N};H)$. In fact, we are interested in a more general setting than that presented in Section~\ref{gensett}. 
In contrast to the scalar-valued case, a new class of subspaces arises in the vector-valued contex, i.e., 
subspaces that are characterized by the pointwise values of a function.

In this section we are going to study this kind of subspaces and their properties. The following condition of localizability has been introduced in~\cite{Vog09}.

\begin{defi}
Let $\mathcal P_{\mathcal W}$ be an orthogonal projection onto a closed subspace $\mathcal W$ of $L^2({{\mathbb R}^N};H)$.
\begin{enumerate}
\item $\mathcal P$  is called \emph{localizable} if
$$\phi f\in{\mathcal W}\qquad\hbox{for all }f\in {\mathcal W}\hbox{ and } \phi\in C^\infty_c({{\mathbb R}^N};{\mathbb C}).$$
\item $\mathcal P$  is called \emph{strictly local} if there exists a family  $(P_x)_{x \in {{\mathbb R}^N}}$ of orthogonal projections onto closed subspaces of $H$ such that $x\mapsto P_x$ is
strongly measurable and such that
\begin{equation}\label{localmeasure}
(\mathcal P f)(x) = P_x(f(x))\qquad \hbox{for a.e.\ }x\in{{\mathbb R}^N}\hbox{ and all }f\in L^2({{\mathbb R}^N};H).
\end{equation}
\end{enumerate}
\end{defi}

\begin{rem}
Observe that the mapping $x\mapsto P_x$ in (2) is essentially bounded, due to contractivity of orthogonal projections.
\end{rem}

Conversely, do all orthogonal projections $\mathcal P$  onto closed subspaces of $L^2({{\mathbb R}^N};H)$ satisfy\eqref{localmeasure}
for some family $(P_x)_{x \in {{\mathbb R}^N}}$ of orthogonal projections onto closed subspaces $G_x$ of $H$?
In general, the above question has a negative answer -- simply think of the scalar-valued radial functions in $L^2(\mathbb R^N)$. 

However, things change if we impose a mild locality assumption. The following result has been proved by H. Vogt in~\cite[Prop.~3]{Vog09}. It essentially relies upon a characterization of local operators proved in~\cite[\S~2]{AreTho05}.

\begin{theo}\label{vogt}
Each localizable orthogonal projection is strictly local.
\end{theo}

\begin{rem}
Conversely, since each orthogonal projection is a contraction, each strongly measurable orthogonal-projection-valued mapping ${{\mathbb R}^N}\ni x\mapsto P_x\in{\mathcal H}$ is essentially bounded and by~\cite[Cor.~2.4]{AreTho05} defines in a canonical way an orthogonal projection onto a closed subspace of $\mathcal H$.
\end{rem}

We address a problem similar to Vogt's. Our aim is to characterize closed ideals of $\mathcal H$, i.e., those closed subspaces $\mathcal I$ of $\mathcal H$ that satisfy the condition\footnote{ This condition only makes sense if $H$ and hence $\mathcal H$  are assumed to have a (complex) Hilbert lattice structure, cf.~\cite[\S~C-I]{Nag86}, what we do throughout.}
$$f\in{\mathcal I}\hbox{ and }|g(x)|\le |f(x)|\hbox{ for a.e.\ }x\in{{\mathbb R}^N}\qquad\hbox{ imply }\qquad g\in{\mathcal I}.$$
We recall that in the theory of lattices, irreducibility of an operator on $\mathcal H$ is the property of leaving invariant no nontrivial closed ideal of $\mathcal H$. 

\begin{rem}\label{omega}
In the scalar case ($H=\mathbb C$) it is well-known that ideals of $L^2({{\mathbb R}^N})$ are exactly those spaces of the form $L^2(\omega)$ for $\omega\subset{{\mathbb R}^N}$, and irreducibility can be easily discussed by means of Ouhabaz's invariance criterium. In particular, it is known that the heat semigroup on $L^p({\mathbb R}^N)$ is irreducible: this follows from the general theory of Sobolev spaces, cf.~\cite[Thm.~4.5]{Ouh05}.

How can the characterization of closed ideals of scalar-valued function spaces be generalized to the vector-valued case? Clearly, this is necessary in order to discuss irreducibility of operators on $\mathcal H$.
\end{rem}

Inspired by the notion of strict localizability, the main problem we address is whether for each closed ideal $\mathcal I$ of $\mathcal H$ there exists a family of closed ideals $(I_x)_{x\in {{\mathbb R}^N}}$ of $H$ such that 
$$
\mathcal I:=\{f \in L^2({{\mathbb R}^N};H): f(x) \in I_x \mbox{ for a.e.\ } x\in  {{\mathbb R}^N} \}.
$$
If we replace the word ``ideal'' by ``subspace'', the answer to this problem is exactly Theorem~\ref{vogt}. In fact, following Vogt's idea we obtain the following.

\begin{theo}
Every orthogonal projection $\mathcal P$ onto a closed ideal is strictly local. The ranges of the orthogonal projections $P_x$ appearing in~\eqref{localmeasure} are in fact closed ideals of $H$ for a.e.\ $x\in{{\mathbb R}^N}$.
\end{theo}

Thus, in the following we may and do identify a projection $\mathcal P$ onto a closed ideal of $\mathcal H$ and a family $(P_x)_{x\in{{\mathbb R}^N}}$ of projections onto closed ideals of $H$.

\begin{proof}
Let us show that $\mathcal P$ is localizable, i.e., that
$$\phi f\in{\mathcal I}\qquad\hbox{for all }f\in {\mathcal I}\hbox{ and } \phi\in C^\infty_c({{\mathbb R}^N};{\mathbb C}).$$
To this end, it suffices to observe that
$$|\phi(x)f(x)|\le \|\phi\|_\infty |f(x)|\quad\hbox{ for a.e.\ }x\in{{\mathbb R}^N}$$
and that $f$ and hence $\|\phi\|_\infty |f|$ belong to $\mathcal I$. It follows by the ideal property that $\phi f\in\mathcal I$.


Then, we can apply Theorem~\ref{vogt} and deduce that ~\eqref{localmeasure} is satisfied by the orthogonal projection onto $\mathcal I$ and by a family  $(P_x)_{x \in {{\mathbb R}^N}}$ of orthogonal projections onto closed subspaces. 

It remains to show that $P_x$ actually projects onto a closed ideal for a.e.\ $x\in{{\mathbb R}^N}$. To this end, recall the following general result: Given an orthogonal projection $Q$ of a Hilbert lattice $X$, its kernel is a closed ideal of $X$ if and only if $|Qy|=Q|y|$ for all $y\in X$. (This easily follows from the comments of~\cite[page 94]{AliBur06} and the fact that each Hilbert lattice is lattice isomorphic to some $L^2(\Theta)$ space, hence by Remark~\ref{omega} its ideals are of the form $L^2(\Xi)$, $\Xi\subset \Theta$).  We are going to check this latter criterion for the kernel of a.e.\ $I-P_x$, i.e., for the range of a.e.\ $P_x$.

Let $f\in L^2({{\mathbb R}^N};H)$. Then 
$$|I-P_x f|(x)=|(I-\mathcal P) f|(x)=(I-\mathcal P)|f|(x)=(I-P_x) |f|(x)$$ 
for a.e.\ $x\in{{\mathbb R}^N}$, i.e., $|(I-P_\cdot) f|=(I-P_\cdot) |f|$. Due to separability of $H$ and by suitable localization arguments, this suffices in order to show that $|(I-P_\cdot) v|=(I-P_\cdot) |v|$ for all $v\in H$.
\end{proof}


Conversely, the following holds.

\begin{prop}
Consider a family  $(P_x)_{x \in {{\mathbb R}^N}}$ of orthogonal projections onto closed ideals of $H$ such that $x\mapsto P_x$ is
strongly measurable. Then the bounded linear operator $\mathcal P$ defined via~\eqref{localmeasure} is an orthogonal projection onto the closed ideal
$$\mathcal I:=\{f\in L^2({{\mathbb R}^N};H):f(x)\in{\rm range}\; P_x\hbox{ for a.e.\ }x\in{{\mathbb R}^N}\}.$$
\end{prop}

\begin{proof}
It follows from H\"older's inequality that $\mathcal P$ is a bounded linear operator on $L^2({{\mathbb R}^N};H)$. The facts that it is a projection, and self-adjoint are consequences of the analogous properties of the operators $P_x$, $x\in{{\mathbb R}^N}$.
\end{proof}

\section{Local symmetries}\label{localsec}

With the aim of generalizing the notion of symmetry considered in the Section~\ref{gensett} to the vector-valued case, it is natural to attach to each point $x\in{{\mathbb R}^N}$ a closed subspace $G_x$ of $H$. We can thus define in analogy to the constant case a subspace
\begin{equation}\label{projgen}
\mathcal{G} := \left\{ f \in L^2({{\mathbb R}^N};H) : f(x) \in G_x \mbox{ for a.\,e. } x\in {{\mathbb R}^N} \right\}
\end{equation}
of $\mathcal H$. Intuitively -- and under suitable smoothness assumptions -- this will define in a natural way a \emph{generalized vector bundle} whose \emph{fibres} are the $G_x$.
Can such a space $\mathcal G$ be left invariant under the time evolution of the diffusion equation? 
This question can be answered by means of Proposition~\ref{gauge}.

More precisely, throughout the remainder of this section we consider an orthogonal-projection-valued mapping $x\mapsto P_x:=P_{G_x}$ to be of class $H^1({{\mathbb R}^N},{\mathcal L}(H))$. Denote by $\mathcal P_{\mathcal G}$ the associated orthogonal projection of $\mathcal H$ onto $\mathcal G$.

\begin{rems}
(1)  Observe  that $P^\perp_\cdot$ is by assumption weakly differentiable and in fact
\begin{equation}\label{nablanegat}
\nabla P_x^\perp=\nabla(Id-P_x)=-\nabla P_x\qquad \hbox{for a.e.\ }x\in{{\mathbb R}^N}. 
\end{equation}
(2) Let us also mention the expression\footnote{ Here and in the following $d(P_x),d(P_x^\perp)$ denote the diagonal $n\times n$-matrices whose diagonal entries all agree with $P_x,P_x^\perp$, respectively.}
\begin{equation}\label{teufel}
\nabla P_x =d(P_x^\perp) (\nabla P_x) P_x + d(P_x) (\nabla P_x) P_x^\perp \qquad \hbox{for  a.e.\ }x\in{{\mathbb R}^N}.
\end{equation}
Formula~\ref{teufel} shows in particular that $\nabla P_x$ (which in general is still a self-adjoint bounded operator but in general \emph{not} a projection) boasts an off-diagonal block structure that is complementary to that of $P_x$. This has been observed in~\cite[(1.15)]{Teu03} in a different context. It keeps its validity (with an analogous proof) in our framework, though.\\
(3) Combining~\eqref{teufel} and $e^{zP_x}=e^z P_x + P_x^\perp$, whose validity for all $z\in{\mathbb C}$ and a.e.\ $x\in{{\mathbb R}^N}$ can be proved as in~\eqref{explform}, one obtains that
\begin{equation}\label{pero}
e^{zd(P_\cdot)}(\nabla P_\cdot)=e^z d(P_\cdot) (\nabla P_\cdot)P^\perp_\cdot + d(P_\cdot^\perp) (\nabla P_\cdot)P_\cdot\qquad\hbox{for all }z\in{\mathbb C}\hbox{ and a.e.\ }x\in{{\mathbb R}^N}.
\end{equation}
\end{rems}

Recall the characterization of closed subspaces of $\mathcal H=L^2({{\mathbb R}^N};H)$ in Theorem~\ref{vogt}.

\begin{prop}\label{neuex}
Let $\mathcal{P}_{\mathcal G}=(P_x)_{x\in{{\mathbb R}^N}}$ be the orthogonal projection onto a closed subspace $\mathcal G$ of $\mathcal H$. Consider the quadratic form $a$  associated with the Laplacian, i.e., 
$$a(f):=\int_{{\mathbb R}^N} \|\nabla f(x)\|^2_{H^N} dx,\qquad f\in H^1({{\mathbb R}^N};H).$$
Then
$$a( e^{is{\mathcal P_{\mathcal G}}}f)=\int_{{\mathbb R}^N} \|(e^{is}-1)e^{-is d(P_x)}(\nabla P_x) f(x)+(\nabla f)(x)\|^2_{H^N} dx.$$
\end{prop}

\begin{proof}
By~\eqref{nablanegat},~\eqref{teufel}, and~\eqref{pero} we can compute
\begin{eqnarray*}
\nabla (e^{is{\mathcal P_{\mathcal G}}}f)(x)&=&\nabla(e^{is}P_x f(x)+ P_x^\perp f(x))\\
&=&e^{is}(\nabla P_x) f(x)+e^{is}d(P_x) (\nabla f)(x) +(\nabla P_x^\perp) f(x)+d(P_x^\perp) (\nabla f)(x)\\
&=& (e^{is}-1)(\nabla P_x) f(x)+e^{is}d(P_x) (\nabla f)(x) +d(P_x^\perp) (\nabla f)(x)\\
&=& (e^{is}-1)(\nabla P_x) f(x)+e^{isd(P_x)} (\nabla f)(x),
\end{eqnarray*}
which holds for all $f\in \mathcal V$ and a.e.\ $x\in{{\mathbb R}^N}$. Accordingly, because $(e^{isP_x})_{s\in\mathbb R}$ and hence $(e^{isd(P_x)})_{s\in\mathbb R}$ are unitary for a.e.\ $x\in{{\mathbb R}^N}$, the form $a$ associated with the Laplacian satisfies
\begin{eqnarray*}
a( e^{is{\mathcal P_{\mathcal G}}}f)&=& \|\nabla (e^{is{\mathcal P_{\mathcal G}}}f)\|^2_{L^2({{\mathbb R}^N};H^N)}\\
&=&\int_{{\mathbb R}^N} \|(e^{is}-1)(\nabla P_x) f(x)+e^{isd(P_x)} (\nabla f)(x)\|^2_{H^N} dx\\
&=&\int_{{\mathbb R}^N} \|(e^{is}-1)e^{-isd(P_x)}(\nabla P_x) f(x)+(\nabla f)(x)\|^2_{H^N} dx=:a_s(f)
\end{eqnarray*}
for all $s\in\mathbb R$ and all $f\in\mathcal V$.
\end{proof}

This shows that in the motivating example introduced in~\eqref{lagrschr},  the Lagrangian ${\mathcal L}(e^{is\mathcal P_{\mathcal G}}\psi)$ stems from a Schr\"odinger equation with  suitable potential that depend on $s$ in a $2\pi$-periodic fashion: in other words, we are led to considering a covariant derivative defined by
$$\nabla_s f:=\nabla f + (e^{is}-1)e^{-isd(P_\cdot)}(\nabla P_\cdot) f$$
where the role of the gauge field is played by the multiplier
$$(e^{is}-1)e^{-isd(P_\cdot)}(\nabla P_\cdot)=(e^{is d(P_\cdot^\perp)}-e^{-isd(P_\cdot)})(\nabla P_\cdot).$$

\begin{rem}
Observe that the above computations bear some formal similarity to the theory of adiabatic perturbation theory, where a somewhat similar magnetic potential is expressed in terms of (time-dependent) spectral projections. Thus, it tempting to interpret the functional
$${\mathcal L}_s(\phi):=\int_0^T |(e^{is}-1)e^{-isd(P_\cdot)}(\nabla P_\cdot)\phi(t)|^2 dt$$
as an interaction Lagrangian and study the mutual asysmptotic behaviour of the time evolution of the systems associated with Lagrangians ${\mathcal L}$ and ${\mathcal L}+{\mathcal L}_s$. However, two related problems soon arise: the form $a_s$ is not symmetric anymore, and accordingly the time evolution of the system associated with ${\mathcal L}+{\mathcal L}_s$ is not governed by a unitary group. In fact, the terms corresponding to the gauge field are $2\pi$-periodic off-diagonal perturbations of the leading term given by the free Hamiltonian $\Delta$ associated with $a=a_0$. 
\end{rem}

By Proposition~\ref{neuex} we are naturally led to introduce the following.

\begin{defi}\label{locconst}
An orthogonal-projection-valued mapping ${{\mathbb R}^N}\ni x\mapsto P_x\in{\mathcal L}(H)$ is called \emph{locally constant} if 
there exists 
a family $(\omega_e)_{e\in \tilde{E}}$ of open disjoint subsets of ${{\mathbb R}^N}$ such that $\mu({{\mathbb R}^N}\setminus \omega)=0$, where $\omega:=\dot{\bigcup}_{e\in E}\; \omega_e$ and 
$P_{|\omega_0}\equiv 0$, $P_{|\omega_\infty}=Id$, and $P_{|\omega_e}$ is the orthogonal projection onto a further closed ideal of $H$.
\end{defi}

In the face of Proposition~\ref{gauge}, we can complement Proposition~\ref{neuex} with the following observation.

\begin{cor}
A necessary condition for the Lie group $(e^{is{\mathcal P}_{\mathcal G}})_{s\in\mathbb R}$ to be a symmetry of the system associated with the Laplacian is that the orthogonal-projection-valued mapping ${{\mathbb R}^N}\ni x\mapsto P_x\in{\mathcal L}(H)$ be locally constant.
\end{cor}

\begin{proof}
It follows by~\eqref{pero} that the gauge field can also be expressed as the multiplier
$$(1-e^{-is})d(P_\cdot) (\nabla P_\cdot)P^\perp_\cdot \oplus (e^{is}-1)d(P_\cdot^\perp) (\nabla P_\cdot)P_\cdot.$$
Such a gauge field vanishes identically, i.e., the covariant derivative satisfies $a_s(f)=a(f)=\| \nabla f\|^2$ for all $s\in\mathbb R$, if and only if the projections $P_\cdot$ are $x$-independent (take e.g. $s=\pi$).
\end{proof}



Let us consider again the setting introduced in Proposition~\ref{neuex}. 

\begin{cor}
Let $A$ be the Laplace operator on $\mathcal H$. The heat semigroup $(e^{tA})_{t\ge 0}$ and hence the unitary group $(e^{itA})_{t\in\mathbb R}$ are irreducible if and only if $H={\mathbb C}$.
\end{cor}

\begin{proof}
If $H=\mathbb C$, the claimed characterization is well-known, cf. Remark~\ref{omega}. If $H\not=\mathbb C$, take a $1$-dimensional closed ideal $I$ of $H$ -- say, the subspace spanned by the first vector of the Hilbert space basis of $H$. Then by Ouhabaz's criterion the closed subspace $L^2({{\mathbb R}^N};I)$ is left invariant under $(e^{tA})_{t\ge 0}$. The claim follows.
\end{proof}

\begin{rem}
Recall that whenever a semigroup $(T(t))_{t\ge 0}$ is positivity preserving (like e.g.\ the semigroup generated by the Laplace operator) on some \textit{scalar}-valued Lebesgue space, its irreducibility is equivalent to saying that positive initial data are instantaneously mapped into strictly positive solutions -- another formulation of the linear heat equation's well known  infinite speed of propagation.


Observe that, even when the heat semigroup is  not irreducibile, for all $f\not\equiv 0$ and all $t>0$ the support
$${\rm supp}\; T(t)f:=\{x\in{{\mathbb R}^N}: T(t)f\not\equiv 0 \hbox{ as an element of } H\}$$ 
is the whole space, just like in the scalar-valued case.
\end{rem}
\begin{rem}
The mapping $x\mapsto P_x$ naturally defines a vector bundle ${{\mathbb R}^N}\to {\mathcal G}\to H$, with $\mathcal G$ as in~\eqref{projgen}. It is not clear to us whether the structure group associated with this vector bundle is related to the usual additive group on $\mathbb R^N$.
\end{rem}

\end{document}